\documentclass[aps,twocolumn,nofootinbib,showpacs,prd,aps,10pt]{revtex4-2}
\usepackage[dvips]{graphicx}
\usepackage[english]{babel}
\usepackage[T1]{fontenc}
\usepackage{mathrsfs}
\usepackage[tbtags]{amsmath}
\usepackage{amssymb}
\usepackage{amsxtra}
\usepackage{xcolor}
\usepackage{amsopn}
\usepackage{latexsym}
\usepackage[mathcal]{eucal}
\usepackage{mathtools}

\newcommand{\BE}{\begin{equation}}
\newcommand{\EE}{\end{equation}}
\newcommand{\BA}{\begin{align}}
\newcommand{\EA}{\end{align}}

\newcommand{\nn}{\nonumber}

\newcommand{\kkEd}{ \frac{{\rm d}^dk_E}{(2\pi)^d}}

\begin{document}

\title{Screened massive expansion of  Schwinger-Dyson equations}

\author{Fabio Siringo}\email{fabio.siringo@ct.infn.it}

\affiliation{Dipartimento di Fisica e Astronomia 
dell'Universit\`a di Catania,\\ 
INFN Sezione di Catania,
Via S.Sofia 64, I-95123 Catania, Italy}

\date{\today}

\begin{abstract}
A general formal derivation of the screened massive expansion is provided by Schwinger-Dyson equations.
Some known issues of the expansion are clarified  and a more general framework is established for a natural extension
of the method to two-loop or to amplitudes which are not directly defined by a generating functional.
For instance,  a one-loop screened expansion is given for the effective gauge-parameter-independent  gluon propagator which arises from the pinch-technique.
\end{abstract}




\maketitle
\newpage

\section{Introduction}

In the last decades, important progresses have been made in the study of the non-perturbative low-energy regime of strong interactions, establishing the dynamical generation of a mass scale in the correlators of the theory
\cite{cornwall,bernard,dono,philip,olive,aguilar04,papa15b,
cucch07,cucch08,cucch08b,cucch09,bogolubsky,
olive09,dudal,binosi12,olive12,burgio15,duarte,aguilar8,aguilar10,aguilar14,papa15,fischer2009,
huber14,huber15g,huber15b,pawlowski10,pawlowski10b,pawlowski13,
varqcd,genself,watson10,watson12,rojas,var,qed,higher,
reinhardt04,reinhardt05,reinhardt14,
dispersion,varT,cucch09b,cucch09c,cucch10,bicudo15,roberts17,
tissier10,tissier11,tissier14,serreau,reinosa,pelaez,pelaez21,2LQ}.

Quite recently, by a mere change of the expansion point, a new perturbative approach has been proposed
for the study of Yang-Mills theory and QCD from first principles in the low-energy  ``non-perturbative''  regime
of strong interactions. A {\it screened massive expansion}  has been 
developed
\cite{ptqcd,ptqcd2,analyt,scaling,xigauge,RG,beta,ghost,damp,thermal,quark,nielsen},
which is perfectly sound in the IR and has many merits of ordinary perturbation theory:
calculability, analytical outputs and a manifest description of the analytic properties in the complex plane.
While the agreement with lattice data is already excellent at one-loop in the gluon sector, some ambiguities on the renormalization 
have been encountered in the full QCD\cite{quark}. 
Moreover, the one-loop contribution to the
quark renormalization function is almost vanishing and 
a two-loop correction would be required\cite{pelaez21,2LQ}.

A two-loop extension of the screened expansion is not straightforward without having first addressed some
minor ambiguities which emerge in its original formulation, like the lack of a rigid criterion for the inclusion of the mass
counterterms in higher-order loops.
On the other hand, there are important physical amplitudes which are not directly defined by a generating functional.
For instance, the pinch-technique 
\cite{aguilar,binosi} 
provides amplitudes with interesting physical features, like being
gauge-parameter independent and fulfilling Abelian Ward identities. Since these amplitudes have an operational
definition from the pinch-technique, it is not obvious how to evaluate them by the screened expansion.

In this paper, we provide a more general framework and formulate the screened expansion as a
loop expansion of the exact Schwinger-Dyson (SD) equations. The formal derivation allows for a straightforward
extension to higher orders and clarifies some unsolved issues of the original expansion. Moreover, in the new framework
the expansion can be easily extended to other theories, 
provided that a specific set of modified SD equations is available.

This paper is organized as follows: the screened expansion is recovered as a loop expansion
of the SD equations in Sec. II; in more detail,  
the simple one-loop approximation is discussed in Sec. IIA, the general higher order extension is studied in Sec. IIB, a specific minimal two-loop expansion is advocated in Sec. IIC
and different truncation strategies are then compared in Sec. IID; in the framework of the pinch-technique,
a screened expansion for the effective gluon propagator is discussed in Sec. III where a simple explicit
calculation at $p=0$ is provided, showing that the effective propagator is finite in the IR;
finally, Sec. IV contains some closing remarks and directions for future work.

\section{Screened expansion of SD equations}

By a loop expansion, the SD equations can be decoupled and are known to provide the standard perturbative
expansion of the vertex functions which define the theory. Here, we introduce a variationally oriented scheme which
captures some non-perturbative features, giving rise to a {\it screened} loop expansion.
We neglect quarks and deal with the pure Yang-Mills theory, but the procedure is general and can be used for the
full QCD or even other theories. 

Suppressing all color and Lorentz indices, the exact SD equations of pure Yang-Mills theory have the following structure
\begin{align}
&\Delta^{-1}=\Delta_0^{-1}-\Pi\nn\\
&\Pi=\Pi\big[\Delta,D,\Gamma_i\big]\nn\\
&D^{-1}=D_0^{-1}-\Sigma\nn\\
&\Sigma=\Sigma\big[\Delta,D,\Gamma_i\big]\nn\\
&\Gamma_i=\Gamma_i\big[\Delta, D,\Gamma_j\big]
\label{SDE}
\end{align}
where all functions have an explicit dependence  on external momenta (not shown) and the vertex functions
$\Sigma$, $\Pi$, $\Gamma_i$ are given functionals of their arguments.
More precisely, in a covariant $R_\xi$ gauge, the gluon propagator is
\BE
\Delta^{\mu\nu}_{ab}(q)=\delta_{ab}\left[ t^{\mu\nu}(q)\Delta(q)-\frac{\xi}{q^2}\,\ell^{\mu\nu}(q)\right]
\label{Delta}
\EE
where $t^{\mu\nu}(q)$, $\ell^{\mu\nu}(q)$ are the projectors
\BE
t^{\mu\nu}(q) =g^{\mu\nu}  - \frac{q^\mu q^\nu}{q^2}\,,\quad
\ell^{\mu\nu}(q) =\frac{q_\mu q_\nu}{q^2}\, ,
\label{tl}
\EE
and $\Delta(q)$ is the transverse part entering in the SD  equations, Eq.(\ref{SDE}), together with its tree-level 
expression
\BE
\Delta_0(q)=-\frac{1}{q^2};
\label{Delta0}
\EE
the gluon self-energy is
transverse
\BE
\Pi^{\mu\nu}_{ab}(q)=\delta_{ab}\, t^{\mu\nu}(q)\, \Pi(q)
\label{Pidef}
\EE
and $\Pi(q)$ is the scalar function entering in Eq.(\ref{SDE}); 
the function $D^{ab}(q)$ is the ghost propagator and $\Sigma_{ab}(q)$ is the ghost self-energy, while the tree-level
ghost propagator is 
\BE
{D_0}^{\,ab}(q)=\frac{\delta_{ab}}{q^2};
\label{D0}
\EE
the set of vertex functions \{$\Gamma_i$\} includes the ghost-gluon vertex $\Gamma_{gh}\equiv\Gamma^{\mu}_{abc}$, 
the three-gluon vertex $\Gamma_3\equiv\Gamma^{\mu\nu\rho}_{abc}$ and
the four-gluon vertex $\Gamma_4\equiv\Gamma^{\mu\nu\rho\sigma}_{abcd}$ which are the only vertices with a 
non-zero tree-level expression $\Gamma_i^{(0)}\not=0$.  The detailed structure of the functionals $\Pi$ and $\Sigma$
is shown in Fig.~1 by diagrams. In the Landau gauge, $\xi=0$, the gluon lines are transverse and are given by the scalar
function $\Delta$. In the general case, each gluon line contains the longitudinal part shown in Eq.(\ref{Delta}), which is exact
and is not affected by the interaction, because the self energy $\Pi$ is transverse.  Thus, in any case, the graphs are 
regarded as functionals of the unknown transverse part $\Delta$.

\begin{figure}[b] \label{fig:SDE}
\vskip 1cm
\centering
\includegraphics[width=0.4\textwidth,angle=0]{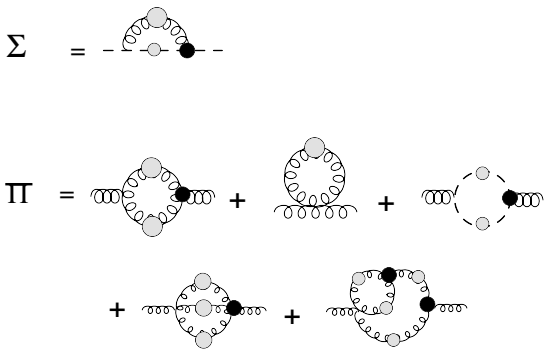}
\caption{The functionals $\Pi$ and $\Sigma$ in the SD equations, Eq.(\ref{SDE}). The black dots are
dressed vertices, while the gray dots label dressed propagators.} 
\end{figure}

Since the vertex functions are expressed as functionals of their arguments, and since there is an infinite
set of vertex functions beyond tree-level, the SD equations cannot be decoupled exactly. 
However, by a loop expansion, the functionals in Eq.(\ref{SDE}) can be expressed as series in powers of the coupling
$g^2=4\pi\alpha_s$, yielding
the standard result of perturbation theory. For instance, at tree-level, the self-energy functionals vanish, $\Sigma=0$, 
$\Pi=0$, while the only non-zero vertices, $\Gamma_{gh}=\Gamma_{gh}^{(0)}$,  $\Gamma_{3}=\Gamma_{3}^{(0)}$,
$\Gamma_{4}=\Gamma_{4}^{(0)}$, are not functionals of other vertices or correlators. The SD equations decouple and 
take the simple form
\begin{align}
\Delta^{-1}&=\Delta_0^{-1}\nn\\
D^{-1}&=D_0^{-1}.
\end{align}
At one-loop, we can replace the propagators and  vertices by their tree-level values inside the loops, and again the SD equations decouple as
\begin{align}
\Delta^{-1}&=\Delta_0^{-1}-\Pi^{(1L)}\big[\Delta_0,D_0,\Gamma_i^{(0)}\big]\nn\\
D^{-1}&=D_0^{-1}-\Sigma^{(1L)}\big[\Delta_0,D_0,\Gamma_i^{(0)}\big]\nn\\
\Gamma_i&=\Gamma_i^{(1L)}\big[\Delta_0, D_0,\Gamma_j^{(0)}\big]
\label{SDE1L}
\end{align}
where the arguments of the one-loop (1L) functionals, on the right hand side, are the known tree-level expressions, yielding 
analytical expressions for the (approximate) propagators and the vertex functions, provided that we are able to evaluate the integrals in the functionals. 

Unfortunately, the standard perturbative expansion breaks down in the infrared and fails
to predict a dynamical generation of mass. In other words, the approximate solutions are very poor at low energy and
become totally unreliable in the limit $q^2\to 0$. We use to say that the dynamical generation of mass is a ``non-perturbative''
effect which cannot be described by a loop-expansion of the exact SD equations at any order. Actually,  it can be shown
that, at any finite order of perturbation theory, by a simple dimensional argument, $\Pi\sim q^2\to 0$ in the limit $q^2\to 0$.
Thus, as it happens in QED, the perturbative gluon propagator still has a pole at $q^2=0$, 
where $\Delta_0^{-1}(0)=0$. 
Without any mass scale available in the original Lagrangian, the perturbative pole
is dictated by the pole of the tree-level propagator $\Delta_0$. Thus, the failure of the perturbative expansion is tightly
linked to the very poor choice of the zeroth-order approximation, since $\Delta_0(q)$ is quite far from the exact solution
of the SD equations in the IR:  the exact solution develops a mass-scale and is finite in the limit $q^2\to 0$ 
\BE
\Delta(0)=\frac{1}{m^2}
\label{zero}
\EE
where $m$ is some dynamically-generated energy whose specific value cannot be predicted by the theory.

The previous analysis suggests that a variationally oriented improvement of the loop expansion could be achieved
by just changing the expansion point and expanding about a massive tree-level propagator
\BE
\Delta_m(q)=\frac{1}{-q^2+m^2}
\label{Deltam}
\EE
which shows the correct limit of Eq.(\ref{zero}) in the IR.

If we look at the first of the one-loop SD equations, Eq.(\ref{SDE1L}),
\BE
\Delta^{-1}-\Delta_0^{-1}=-\Pi\sim{\cal O}(g^2)
\EE
we get a mismatch in the IR, with the difference on the left-hand side which gets close to $m^2$ and the right-hand side which must be
a small perturbative correction. It is quite obvious that the perturbative expansion must break down since the difference
is not a  small correction. While, if we attempted an expansion about $\Delta_m$, the difference $\Delta-\Delta_m$ would
be small at any energy scale and would vanish in the IR: the perturbative correction would be very small and we could extract
reliable approximations from the exact SD equations, already at the one-loop level.

We observe that, even if the bare propagator $\Delta_0$ appears in the SD equations, the exact solution $\Delta$
does not need to be close to $\Delta_0$ (and is not). We can rearrange the SD equations and eliminate $\Delta_0$ by
using the exact relation
\BE
\Delta_0^{-1}=\Delta_m^{-1}-m^2.
\EE
The first pair of exact SD equations can be recast as
\begin{align}
\Delta^{-1}&=\Delta_m^{-1}-\Pi^\prime\nn\\
\Pi^\prime&=m^2+\Pi\big[\Delta,D,\Gamma_i\big].
\label{SDprime}
\end{align}
It is quite obvious that the change has no effect on the exact solution of the equations which does not depend on the
parameter $m$. We can use $m$ as a variational parameter and optmize the expansion by requiring that
$\Delta_m\approx \Delta$. In that sense, the parameter $m^2$ does need to be exactly the same scale $1/\Delta(0)$
encountered in Eq.(\ref{zero}), but should be tuned in order to optimize the expansion.
We will be back to the point later. At the moment, let us just suppose that we picked up the
best value and the difference between the propagators is ``small'':
\BE
\Delta^{-1}-\Delta_m^{-1} =-\lambda\Pi^\prime \approx 0
\label{lambda}
\EE
where $\lambda$ is just a fictitious expansion parameter to be set to 1 at the end. We might attempt a double
expansion: a {\it $\delta$-expansion} in powers of $\lambda$ and a loop expansion in powers of the coupling $g^2$. 
If the coupling $g^2$ is not too large,
and we have now evidence that it is not\cite{ptqcd,ptqcd2,RG}, 
the optimized loop expansion gives reliable predictions, even in the IR,
{\it provided} that Eq.(\ref{lambda}) is fulfilled, i.e. that we are expanding about the optimal point.
As long as $\Pi^\prime$ is small, even if $\Pi$ takes a moderate value, the expansion makes sense. Then,
in the first place, we assume that Eq.(\ref{lambda}) is satisfied and that we might truncate 
the expansion at some low order in $\lambda$.

Starting from Eq.(\ref{lambda}), the expansion of $\Delta$ in powers of $\lambda$ yields
\BE
\Delta=\frac{1}{\Delta_m^{-1}-\lambda\Pi^\prime}=\Delta_m+\lambda\Delta_m \Pi^\prime \Delta_m+{\cal O}(\lambda^2)
\label{exp1}
\EE
and, at first order, the variation $\delta\Delta=\Delta-\Delta_m$ reads
\BE
\delta \Delta=\lambda  m^2 \left(\Delta_m\right)^2+\lambda\left(\Delta_m\right)^2 \,\Pi\big[\Delta_m,D,\Gamma_i\big]
\label{dDelta}
\EE
where we have used the second of the  exact SD equations, as given by Eq.(\ref{SDprime}), and
$\Pi$ is evaluated at $\lambda=0$. We observe that the exact propagator has disappeared in the variation $\delta \Delta$
which is expressed in terms of $\Delta_m$, but still depends on the exact functions $D$ and $\Gamma_i$.
Corrections of order $\lambda^2$, or higher order, can be introduced by the same method as required.

Using the variation $\delta \Delta$, Eq.(\ref{dDelta}), we can evaluate the first variation of the SD equations and
eliminate the dependence on the exact $\Delta$ among the arguments of the functionals. In detail, the expansion
of Eq.(\ref{SDE}) leads to the approximate first-order SD equations  
\begin{align}
&\Delta^{-1}=\Delta_m^{-1}-\Pi^\prime\nn\\
&\Pi^\prime=m^2+\Pi\big[\Delta_m, D,\Gamma_i\big]+\int dk \>
\left(\frac{\delta \Pi}{\delta\Delta(k)}\right)_{\Delta_m} \!\!\!\delta \Delta(k)\nn\\
&D^{-1}=D_0^{-1}-\Sigma\nn\\
&\Sigma=\Sigma\big[\Delta_m,D,\Gamma_i\big]+\int dk \>
\left(\frac{\delta \Sigma}{\delta\Delta(k)}\right)_{\Delta_m} \!\!\!\delta \Delta(k)\nn\\
&\Gamma_i=\Gamma_i\big[\Delta_m, D,\Gamma_j\big]+\int dk \>
\left(\frac{\delta \Gamma_i}{\delta\Delta(k)}\right)_{\Delta_m} \!\!\!\delta \Delta(k)
\label{SDE1st}
\end{align}
where $dk$ is the appropriate integration measure. According to its definition in Eq.(\ref{Pidef}), here and in all the following equations, the functional $\Pi$ is the transverse projection of the graphs which define it. While individual graphs might
contain a longitudinal part, that part does not give any contribution to the transverse component $\Delta$ 
of the gluon propagator in the SD equations. On the other hand, the exact resummation of all the longitudinal parts must vanish because the exact $\Pi$ is transverse.

The first-order equations are not decoupled yet, because we still have the exact (unknown) functions 
$D$ and $\Gamma_i$ 
among the arguments of the functionals on the right hand side.
As we did for the standard one-loop expansion, in Eq.(\ref{SDE1L}),
we can get explicit decoupled equations by a loop expansion in powers of the coupling $g^2$.

\subsection{First order, one loop expansion}

If only one loop graphs are retained in the expansion of the functionals in the first-order SD equations, Eq.(\ref{SDE1st}),
we have some important simplifications. First of all, inside the loops we can drop the last term of $\delta \Delta$ in 
Eq.(\ref{dDelta}), which already is of order $g^2$ and would give rise to higher order terms. Thus we can just set
\BE
\delta \Delta=(\delta \Delta)_0=\lambda  m^2 \left(\Delta_m\right)^2.
\label{dDelta1L}
\EE
Moreover, for any generic functional ${\cal F}[\Delta]$ which does not depend on $m^2$, we can write 
\begin{align}
\frac{\partial}{\partial m^2}{\cal F}\big[\Delta_m\big]&=\int dk \>
\left(\frac{\delta {\cal F} }{\delta\Delta(k)}\right)_{\Delta_m} \!\!\!\ \frac{\partial \Delta_m(k)}{\partial m^2}\nn\\
&=-\int dk \>
\left(\frac{\delta {\cal F} }{\delta\Delta(k)}\right)_{\Delta_m} \!\!\!\ \left[\Delta_m(k)\right]^2
\end{align}
and obtain the identity
\BE
\int dk \>
\left(\frac{\delta {\cal F} }{\delta\Delta(k)}\right)_{\Delta_m} \!\!\!\ \left(\delta \Delta(k)\right)_0=
-\lambda m^2\frac{\partial}{\partial m^2}{\cal F}\big[\Delta_m\big].
\label{id}
\EE
Finally, at one-loop, we can replace the ghost propagator and the vertices by their tree-level 
functions inside the loops and, having set $\lambda=1$, we obtain the decoupled first order SD equations
\begin{align}
&\Delta^{-1}=\Delta_m^{-1}-\Pi^\prime\nn\\
&\Pi^\prime=m^2+\left(1-m^2\frac{\partial}{\partial m^2}\right)
\Pi^{(1L)}\big[\Delta_m, D_0,\Gamma_i^{(0)}\big]\nn\\
&D^{-1}=D_0^{-1}-\Sigma\nn\\
&\Sigma=\left(1-m^2\frac{\partial}{\partial m^2}\right)
\Sigma^{(1L)}\big[\Delta_m,D_0,\Gamma_i^{(0)}\big]\nn\\
&\Gamma_i=\left(1-m^2\frac{\partial}{\partial m^2}\right)\Gamma_i^{(1L)}\big[\Delta_m, D_0,\Gamma_j^{(0)}\big].
\label{SDE1st1L}
\end{align}
The explicit graphs contributing to $\Pi^\prime$ and $\Sigma$ are shown in Fig.~2, where the {\it crossed graphs} contain a transverse mass counterterm $m^2 t^{\mu\nu}$, shown as a cross, originating from the mass derivative of the standard graphs and taking in  account
the variation $\Delta_m\to \Delta_m+\Delta_m m^2\Delta_m$ inside the loops. Of course, the functional $\Pi^{(1L)}$
is the transverse projection of the graphs in Fig.~2, since the  graphs contain a longitudinal part.

\begin{figure}[b] \label{fig:1loop}
\vskip 1cm
\centering
\includegraphics[width=0.22\textwidth,angle=-90]{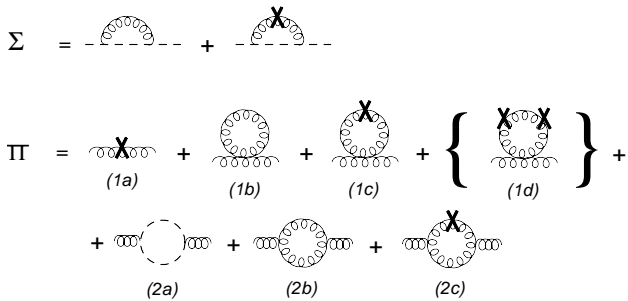}
\caption{Graphs contributing to the one-loop screened expansion of $\Pi^\prime$ and $\Sigma$.
The doubly crossed gluon tadpole, graph (1d) in braces, is not present in the minimal expansion.}
\end{figure}

The resulting expansion is almost equivalent to the massive expansion introduced 
in Refs.\cite{ptqcd,ptqcd2,analyt,scaling,xigauge}. 
The only difference is the absence of the doubly crossed tadpole
which was included in the previous works and is also shown in braces in Fig.~2. 
Here, at first order in $\lambda$, there are no doubly crossed graphs, but they would
be included by adding higher order corrections.
That missing graph is constant and finite and its absence can be absorbed in part by
a shift of the parameter $m$ and other renormalization constants, as discussed in Ref.\cite{ptqcd2}.
It must be kept in mind that, because of the truncation, ambiguities like that can arise, especially regarding finite
higher order contributions. It is not obvious why the result should be improved by their further inclusion but, somehow,
these terms might mimic the effect of higher loops and their inclusion could
improve the agreement with the exact result. In fact, the doubly crossed tadpole in Fig.~2, which arose by a strict vertex
counting in the original screened expansion\cite{ptqcd,ptqcd2}, introduces a slight improvement and leads to an excellent agreement with the lattice data\cite{xigauge,beta}.
Moreover, in some different frameworks, like the modified SD equations of the pinch-technique, the existence of that
term would become crucial, as shown in Sec.~\ref{PT}.  We will discuss more general and improved truncation strategies below, in Sec. II-D.
Despite that shortcoming, the present minimal one-loop expansion has the advantage of a straight derivation from the SD equations and a well defined extension to higher orders. Besides,
we expext that the ambiguities on the truncation would become less relevant when higher loop corrections are added in the expansion. 

We observe that the added mass scale breaks the Becchi-Rouet-Stora-Tyutin (BRST)  symmetry at any finite
order of the expansion, which is not protected from the appearance of new spurious diverging terms with
dimensions of $m^2$. While in the UV the mass parameter becomes irrelevant, and the usual diverging terms
are absorbed by the standard counterterms, a diverging mass term cannot be canceled because there are no
mass parameters in the original Lagrangian. However, since we just rearranged the expansion of the exact SD equations,
the spurious divergences must cancel somehow.  In fact, the differential operator $(1-m^2\partial/\partial m^2)$ does
the job and cancels all spurious mass divergences in the expansion. In dimensional regularization, taking $d=4-2\epsilon$,
spurious terms of the kind $\sim m^2/\epsilon$, are found in most of the graphs of Fig.~2  for the self energy
$\Pi^{(1L)}$. All these spurious terms are canceled since
\BE
\left(1-m^2\frac{\partial}{\partial m^2}\right) \left[\frac {m^2}{\epsilon}\right]=0.
\label{divcanc}
\EE
The graphs come in pairs, with each loop accompanied by the corresponding crossed loop, and the spurious
divergence cancels in their sum according to Eq.(\ref{divcanc}).
Thus the one-loop expansion of Eq.(\ref{SDE1st1L}) shows the same identical diverging terms of the standard
loop expansion and can be renormalized by the standard set of counterterms\cite{ptqcd,ptqcd2,analyt}.

An important point is that, having inserted an arbitrary mass, there are two energy scales in the quantized theory:
the renormalization point $p=\mu$ which comes from the regularization of the loops and the mass parameter $m$.
Thus, even if the calculation is from first principles and the exact SD equations derive from the full, gauge fixed, 
Faddeev-Popov Lagrangian of QCD, the expansion contains one free parameter which can be taken as the ratio $m/\mu$.
The expansion can be {\it optimized} by a variational choice of the best parameter. For instance, it was shown that
enforcing the gauge invariance of poles and residues provides 
an excellent agreement with the lattice data\cite{xigauge}.

\subsection{Higher orders and loops}
In principle, we could extend the expansion and truncate it at the order $\lambda^N$ and $(g^2)^L$ which we call
$N$th-order and $L$-loop. But, first of all, we observe that it does not make sense to consider a large value of $N$ at 
a small value of $L$. While the $\lambda^N$ term is accompanied by a power of $(\Pi^\prime)^N=(m^2+\Pi)^N$,
the term $\Pi^N$ has no effects inside the loops, even in one-loop graphs, if $N>L-1$, since the generated graphs
would have $N+1>L$ loops at least and would be discarded in the $L$-loop expansion. 
We would just add powers of $m^2$, which are not regarded as 
small terms according to Eq.(\ref{lambda}). In fact, even for $N=1$ the self energy $\Pi$ does not contribute
to $\delta \Delta$ in Eq.(\ref{dDelta1L}) at one-loop. Thus, there is no reason to believe that the result might
improve by increasing $N$.
For instance, at one loop, increasing $N$ would only produce a proliferating of crossed graphs, like the doubly crossed
tadpole of the original screened expansion\cite{ptqcd,ptqcd2}. These terms introduce finite small corrections which
might improve the result sometimes, but it is not
obvious why they should in general.

However, the crossed terms, which cancel the spurious 
divergences, originate right from the insertions of $m^2$ which play an important role: the spurious divergences are introduced
by the mass scale $m$ and can only be canceled by those insertions of $m^2$.  Moreover, we know that the divergences must cancel
exactly if the whole series is resummed, as BRST symmetry would be restored. Then, we must allow for a sufficient
number of  mass insertions in the graphs and an order $N>1$ is required beyond one-loop.
On the other hand, according to Eq.(\ref{exp1}), 
any power of $\lambda$ adds at least one loop to the graphs, so that some $L$-loop
terms would be missing in the expansion if $N<L$. Then, we advocate the choice of $N=L$ as the {\it minimal} compromise
which might cancel all spurious divergences without a large proliferation of crossed graphs.

At a generic order $N$, the simple procedure which we described above becomes quite cumbersome and a direct
iteration of Dyson equations would be easier to be implemented in practice. But it is instructive to
look at the details anyway.
The variation $\delta \Delta$ of Eq.(\ref{dDelta})  would generalize as 
\BE
\delta \Delta=\Delta_m\sum_{n=1}^N\left[ \left(m^2+\Pi\right) \Delta_m\right]^n\lambda^n
\label{dDeltaN}
\EE
where $\Pi$ is a functional $\Pi=\Pi\big[\Delta,D,\Gamma\big]$ which contains the original exact arguments.
Thus,  a recursive iteration is understood, by insertion of  Eq.(\ref{dDeltaN})
and  Eq.(\ref{SDE})  for $\Sigma$ and $\Gamma_i$ in the functionals, up to the desired power orders $N$ and $L$.
Moreover, all functionals in the SD equations should be expanded according to the generic Taylor series
\begin{align}
&{\cal F}\big[\Delta\big]={\cal F}\big[\Delta_m\big]+\nn\\
&+\sum_{n=1}^{N}
\frac{1}{n!}\int
\left(\frac{\delta^n {\cal F} } {\delta\Delta(k_1)\dots \delta\Delta(k_n)}\right)
\prod_{i=1}^n\left[\delta \Delta(k_i)dk_i\right]
\label{SDEN}
\end{align}
until the desired order $\lambda^N$ and $(g^2)^L$ is reached, discarding higher order terms.

As remarked above, for $N>1$, it is  easier to follow the straightforward path of iterating the Dyson equations directly. 
The SD equations
can be written in the exact equivalent form
\begin{align}
&\Delta=\Delta_m+\Delta_m\>\lambda\left(m^2+\Pi\right)\Delta\nn\\
&D=D_0+D_0\,\Sigma\, D\nn\\
&\Pi=\Pi\big[\Delta,D,\Gamma_i\big]\nn\\
&\Sigma=\Sigma\big[\Delta,D,\Gamma_i\big]\nn\\
&\Gamma_i=\Gamma_i\big[\Delta, D,\Gamma_j\big]
\label{SDEdyson}
\end{align}
where $\lambda$ must be set to 1 at the end of the calculation.
The equations can be iterated, up to the desired order, discarding all terms with higher powers than $\lambda^N$ and
$(g^2)^L$.  The procedure will generate exactly the same graphs as described before. For instance, it is easy to check
that for $L=1$ and $N=1$ we obtain the same graphs of Fig.~2. Both  procedures are exact expansion of the same set of
SD equations, then at a given order they must give the same set of graphs.

An even easier procedure arises by a two step expansion of the SD equations. Observing that
\begin{align}
\Delta^{-1}&=\Delta_0^{-1}-\lambda\Pi\nn\\
\Delta_0^{-1}&=\Delta_m^{-1}-\lambda m^2
\label{2step}
\end{align}
we can iterate the standard Dyson equation, equivalent to the first line of Eq.(\ref{2step}),
\BE
\Delta=\Delta_0+\Delta_0 (\lambda \Pi) \Delta
\label{dyson1}
\EE
and generate the usual loop expansion of the SD equations. At each iteration this equation adds one or more loops and
a power of $\lambda$. Then, as a second step, we
can iterate the second Dyson equation, equivalent to the second line of Eq.(\ref{2step}),
\BE
\Delta_0=\Delta_m+\Delta_m (\lambda m^2) \Delta_0
\label{dyson2}
\EE
which inserts mass counterterms in the ordinary graphs and replaces $\Delta_0$ by $\Delta_m$. 
Each iteration adds a cross in a gluon line and a power of
$\lambda$.
For instance, starting from a one-loop term of $\Pi$, and inserting one loop by Eq.(\ref{dyson1}), we can still add
$N-1$ crosses at the order $N$.
At any given loop-order, all graphs are readily obtained by just adding the allowed number of crosses to the gluon lines.
In practice, just draw all the ordinary graphs of perturbation theory and decorates them inserting the correct number
of crosses in all the topologically different positions.

\subsection{Minimal two loop extension}

Let us describe a minimal two-loop expansion with $N=L=2$. By the two-step procedure described in 
the previous subsection, we first iterate Eq.(\ref{dyson1}) inside the SD equations and generate the standard two-loop
expansion. There are three classes of graphs contributing to the  gluon self-energy $\Pi[\Delta_0]$ at the order $L=2$:

i) a first class $\Pi^{(1L)}_0[\Delta_0]$ is given by the one-loop graphs of Fig.~1, with the exact propagator $\Delta$ replaced 
by $\Delta_0$, according to Eq.(\ref{dyson1}) at the order $N=0$,   
and the other arguments $D$, $\Gamma_i$ set to their bare values (these graphs have no powers of $\lambda$, 
then $N=0$ and $L=1$);

ii) a second class $\Pi^{(2L)}_1[\Delta_0]$ originates from the insertion of $\lambda \Pi$ in the one-loop graphs of Fig.~1, according to Eq.(\ref{dyson1}) at the order $N=1$, with all the arguments set to their bare values 
(these graphs have a power of $\lambda$ and we retain only two-loop terms, then $N=1$ and $L=2$);

iii) a third class $\Pi^{(2L)}_0[\Delta_0]$ is given by the two-loop graphs of Fig.~1 where $\Delta$ is just replaced by
$\Delta_0$ and all the other arguments are also set to their bare values (any further insertion would
raise the loop order beyond two-loop, then $N=0$ and $L=2$ for these graphs).

There would be a fourth class of graphs, generated by the one-loop graphs of Fig.~1 by iterating with the ghost and vertex
functionals of Eq.(\ref{SDEdyson}): for instance, inserting the
ghost propagator in a one-loop graph
\BE
D=D_0+D_0\,\Sigma\, D_0+\cdots
\EE
the graph generates a two-loop graph with $N=0$. However, these two-loop graphs can be added to the third class
$\Pi^{(2L)}_0[\Delta_0]$. Then, all graphs in the three classes are understood as functionals of
the bare arguments, i.e. all internal lines and vertices are replaced by the bare ones (in the notation, we are omitting the other bare arguments, $D_0$, $\Gamma^{(0)}_i$, for brevity).

In the second step, according to Eq.(\ref{dyson2}), we decorate the graphs with the allowed number of mass insertions
and replace all the bare gluon lines $\Delta_0$ by the massive ones, $\Delta_m$.

All graphs in the second class $\Pi^{(2L)}_1[\Delta_0]$ have $N=1$ and can receive only one further mass insertion. 
The generated graphs can be drawn by just inserting one mass counterterm in a gluon line in all the  
different positions.
Following the same 
steps that led to Eq.(\ref{SDE1st1L}),  the total set of graphs generated by the second class can be written as
\BE
\Pi^{(ii)}=\left(1-m^2\frac{\partial}{\partial m^2}\right)
\Pi_1^{(2L)}\big[\Delta_m \big].
\label{Pi2}
\EE
The first class $\Pi^{(1L)}_0[\Delta_0]$ and the third class $\Pi^{(2L)}_0[\Delta_0]$ contain graphs with $N=0$ which
can receive a maximum of two mass insertions. The total set of graphs generated, including the correct combinatorial factors, can be written as
\begin{align}
\Pi^{(i)}&=\left(1-m^2\frac{\partial}{\partial m^2}  +\frac{1}{2} m^4 \frac{\partial^2}{\partial (m^2)^2}\right)
\Pi_0^{(1L)}\big[\Delta_m \big]\nn\\
\Pi^{(iii)}&=\left(1-m^2\frac{\partial}{\partial m^2}  +\frac{1}{2} m^4 \frac{\partial^2}{\partial (m^2)^2}\right)
\Pi_0^{(2L)}\big[\Delta_m \big].
\label{Pi1}
\end{align}
Thus, the evaluation of the generated graphs is immediate if the analytical expressions are known for the
standard two-loop graphs with massive gluon lines (all of them occur in the Curci Ferrari model\cite{gracey2L}).
The proof of Eq.(\ref{Pi2}) and Eq.(\ref{Pi1}) relies on the identity
\BE
\Delta_m^2=-\frac{\partial}{\partial m^2}\Delta_m
\EE
which can be used recursively if the functionals depend on $m^2$ only through the argument $\Delta_m$.
Actually, each graph in $\Pi\big[\Delta_m\big]$  contains a product of massive gluon propagators 
$\Delta_i\equiv\Delta_m(k_i)$,
with $i=1,2\dots M$ where $M$ is the number of internal gluon lines. The derivative of this product reads
\BE
\frac{\partial}{\partial m^2} \prod_{i=1}^M \Delta_i=\sum_{j=1}^M (-\Delta_j^2)\prod_{i\not=j}\Delta_i.
\EE
Then, multiplying by $m^2$, the differential operator $-m^2\partial/\partial m^2$ replaces a transverse
gluon line by the chain $\Delta_j m^2\Delta_j$ in all the different positions in the graph, recovering Eq.(\ref{Pi2}).
A second derivative gives
\begin{align}
\frac{\partial^2}{\partial (m^2)^2} \prod_{i=1}^M \Delta_i&=\sum_{j=1}^M (2\Delta_j^3)\prod_{i\not=j}\Delta_i\nn\\
&+\sum_{j=1}^M (\Delta_j^2) \sum_{k\not=j} (\Delta_k^2)\prod_{i\not=j,k}\Delta_i
\end{align}
then multiplying by $m^4/2$
\begin{align}
\frac{m^4}{2}\frac{\partial^2}{\partial (m^2)^2} \prod_{i=1}^M \Delta_i&
=\sum_{j=1}^M (\Delta_j m^2\Delta_j m^2\Delta_j)\prod_{i\not=j}\Delta_i\nn\\
+\sum_{j<k} &(\Delta_j m^2 \Delta_j) (\Delta_k m^2\Delta_k)\prod_{i\not=j,k}\Delta_i
\label{m4}
\end{align}
where each pair $j,k$ is taken only once in the last line. We recognize a double mass insertion in the same line $j$,
summed over all the different gluon lines, in the first term. In the second term we find 
two mass insertions in different lines $j,k$, summed over all the pairs, each taken once. 

\begin{figure}[t] \label{fig:3 check}
\vskip 1cm
\centering
\includegraphics[width=0.35\textwidth,angle=-90]{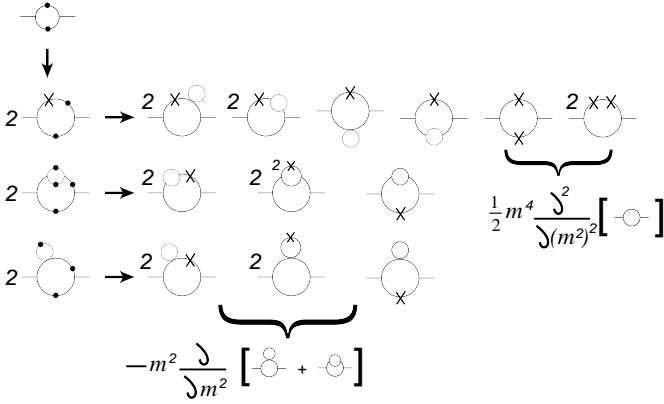}
\caption{All graphs generated by the first one-loop graph of Fig.~1 after a double iteration of the SD equations,
according to Eq.(\ref{SDEdyson}), limiting to pure gluon terms, neglecting any ghost and vertex insertions for brevity. 
All straight lines are massive gluon propagators while the black dots label exact, dressed, propagators.  Only dressed
propagators can receive further insertions in the next iteration.}
\end{figure}

It is instructive to see how the same graphs are generated by a direct recursive iteration of Eq.(\ref{SDEdyson}),
as shown in Fig.~3 for a single one-loop graph. 

The present two-loop expansion produces a really
minimal set of graphs, containing all the standard two-loop graphs and a minimal set of crossed graphs which
cancel the spurious divergences. Any proliferation of crossed graphs is avoided, and still the expansion
might be renormalized, in principle, by the standard set of counterterms and renormalization constants of perturbation theory. In other words, all spurious diverging terms, which cannot be canceled by the counterterms, are removed by
the differential operators in Eqs.(\ref{Pi2}),(\ref{Pi1}).

In order to illustrate the mechanism, we observe that all diverging graphs can be expanded to first order in
external momenta and reduced to scalar diverging functions\cite{misiak2L,gracey2L}.
We are interested in the spurious diverging terms of the self energy $\Pi(p)$ in the limit $p^2\to 0$, which
are not canceled by the wave function renormalization constant at any order. The massive gluon lines give rise to
a diverging result $\Pi(0)=c\times m^2$ where the constant $c$ contains poles $1/\epsilon^n$ and logarithmic terms
$(\log m)^\ell/\epsilon^n$. 
At two loop, we would be left with  dangerous double poles $m^2/\epsilon^2$ and logarithmic poles
$m^2\log m/\epsilon$ with dimensions of $m^2$. For instance, at $p=0$, the diverging part of a typical two-loop graph
is given by the Euclidean integral\cite{ford2L,gracey2L}
\begin{align}
&\left(\frac{g^2\mu^{2\epsilon} }{(2\pi)^d}\right)^2\int\, 
\frac{{\rm d}^{4-2\epsilon} k\>{\rm d}^{4-2\epsilon} q }
{(k^2+m^2)[(k-q)^2+m^2)](q^2+m^2)}\nn\\
&=-\frac{3g^4}{32\pi^2}\left[\frac{m^2}{\epsilon^2}+\frac{3m^2}{\epsilon}
+2\frac{m^2}{\epsilon}\log({\overline \mu}^2/m^2)+\dots\right].
\label{typ2}
\end{align}
It is quite obvious that the simple pole and the double pole $m^2/\epsilon^2$ 
disappear after the action of the differential operators in
Eqs.(\ref{Pi2}),(\ref{Pi1}), while the logarithmic pole is reduced to a simple pole
\begin{align}
\left(1-m^2\frac{\partial}{\partial m^2}\right) \frac{m^2\log (\overline\mu^2/m^2)}{\epsilon}&=\frac{m^2}{\epsilon}\nn\\
\left(1-m^2\frac{\partial}{\partial m^2}  +\frac{1}{2} m^4 \frac{\partial^2}{\partial (m^2)^2}\right)
\frac{m^2\log (\overline\mu^2/m^2)}{\epsilon}&=\frac{m^2}{2\epsilon}.
\label{logm}
\end{align}
The residual simple poles are what one would expect since, at two loop,  a renormalization of the coupling 
has to be considered in the one loop graphs, yielding the same kind of divergence.  In the limit $p^2\to 0$, 
a typical one-loop graph
contains the spurious diverging part
\BE
\Pi(0)\sim g^2 m^2\left[\frac{1}{\epsilon}+\log(\mu^2/m^2)+C\right]
\EE
which is canceled by the crossed graph, according to Eq.(\ref{divcanc}), leaving
\BE
\Pi(0)\sim g^2 m^2.
\EE
At two-loop order, we must consider the renormalization of the coupling constant up to one loop
\BE
g^2=g^2_R Z_g^2=g^2_R(1+2\delta Z_g)
\EE
where, in the $\overline{\text{MS}}$ scheme, $\delta Z_g= b g^2_R/\epsilon$, with a  constant $b$ to be defined 
in order to cancel the divergence of the two-loop graphs.
Then, the finite sum of the one-loop
graphs acquires an extra two-loop diverging term
\BE
\Pi(0)\sim g^2m^2= g_R^2 m^2+2b g^4_R\frac{m^2}{\epsilon}
\EE
which, in principle, could subtract the residual divergences arising from  Eqs.(\ref{logm}),(\ref{typ2}), without having to add
any mass counterterm by hand.
Of course, a detailed two-loop calculation is required in order to check that the residual UV divergences can
be absorbed by the same set of counterterms of the standard perturbation theory, as we expect since the mass parameter becomes irrelevant at high energy.
 
\subsection{Different truncation strategies}

As discussed above, an ambiguity can arise about the number of finite crossed graphs to be retained
at any loop order of the expansion.  Different truncation strategies might lead to slightly different results, especially
at one-loop. In this section we compare the minimal expansion with the more effective vertex-counting criterion of the original derivation in Refs.\cite{ptqcd,ptqcd2}.

First of all, we observe that a strict loop-wise expansion would lead to the same identical result of 
standard perturbation theory. Since $m^2$ insertions do not change the loop order, in principle, at any given loop
order, we should resum all graphs with any number of mass insertions. The exact resummation of all the mass insertions
\BE
\Delta_m+\Delta_m m^2 \Delta_m+\Delta_m m^2\Delta_m m^2\Delta_m\dots=\Delta_0
\label{resum}
\EE
is equivalent to restoring $\Delta_0$ in place of $\Delta_m$ in any internal gluon line.
On the other hand, the infinite sum in Eq.(\ref{resum})  provides some non-perturbative content which makes the
difference between a screened expansion about $\Delta_m$ and the standard perturbative expansion about $\Delta_0$.
Thus, the above series in Eq.(\ref{resum}) {\it must} be truncated, since its exact sum would wash out the non-perturbative
content. Retaining a finite number of terms in any gluon line would lead to {\it crossed} graphs, all belonging to the same
loop order. Especially when these graphs are finite, their inclusion does not change the result dramatically, and different
strategies might be envisaged for determining their inclusion at any loop order.

The minimal choice $N=L$ has been shown to be effective for canceling all spurious divergences, but it arose from
the assumption that $\lambda \Pi^\prime=\lambda(m^2+\Pi)$ is  a {\it small} quantity, validating the $\delta$-expansion in powers of $\lambda$. However, in the loop expansion, the exact function $\Pi$ is replaced by its  truncated expansion
and even a finite missing contribution might lead to a large  $\Pi^\prime$, thus invalidating the expansion.
A notable example is provided by the modified SD equations arising from  the pinch-technique, since the one-loop
effective self-energy $\hat \Pi^{(1L)}(0)=0$ exactly, because of the QED Ward identity which is satisfied by the modified
vertices. 
The anomaly arises because the expansion {\it parameter} is by itself a truncated  expansion.

Even if more cumbersome, a  vertex-counting criterion would be more reliable, as it is based on a $\delta$-expansion
in powers of the whole interaction, which does not depend on the loop order of the expansion.
When starting from a well defined Lagrangian, perturbation theory leads to an expansion of the generating functional
in powers of the whole interaction. In the resulting graphs, each power of  a local interaction term introduces a vertex in the expansion. In presence of anomalous interaction terms, which do not contain powers of the
coupling $g^2$,  we can only take track of the order by just counting the number of vertices in a graph. 
Thus, regarding the mass counterterm insertions as two-point vertices, it makes sense to determine the order of
a graph by counting the total number of vertices. This {\it democratic} criterion gives the same order to $m^2$
counterterms, three-gluon and four-gluon vertices, which would be of order $g^0$, $g$ and $g^2$, respectively, 
in a loop-wise expansion. Actually, we can check that all graphs in Fig.~2, including the doubly crossed tadpole,
have no more than three vertices.

Implementing the same  procedure for the expansion of the SD equations is not immediate without going through
the functional definition of the theory. However, we can {\it assume} that a given set of SD equations  
derives from some unknown functional definition of the theory and that the expansion of the SD equations by graphs
can be traced back to a power expansion of the whole interaction. Then, by the democratic criterion, we can truncate
the screened expansion and retain graphs with a given number of vertices. At one-loop, in order to cancel all the
spurious divergences we should retain graphs with no more than three vertices. 
Especially at one-loop, where the ambiguity on the retained terms can have a stronger effect, the vertex-counting
criterion is more reliable, as shown by the agreement which is found with the lattice data. In fact, by a variational
argument, when expanding about $\Delta_m$, if the massive propagator $\Delta_m$ is 
a very good approximation for the exact propagator $\Delta$, then
the effect of the whole interaction must be very small and a perturbative expansion in powers of the whole interaction,
including the mass counterterm, makes sense.

\section{Screened expansion and pinch-technique}
\label{PT}

The pinch-technique\cite{binosi,aguilar} is a general method based on a new set  of off-shell
Green's functions which are independent of the gauge-fixing
parameter and satisfy ghost-free Ward identities.  Historically, the method
was first introduced as a tool for the study of the SD equations\cite{cornwall,cornwall89}.
Actually, the arguments of the SD equations, Eq.(\ref{SDE}), are  gauge-dependent {\it unphysical} 
Green’s functions, while all {\it physical} observable must be gauge-parameter-independent.
The delicate all-order cancellation of the gauge dependence might be distorted by an arbitrary
truncation of the infinite set of equations. On the other hand, in the pinch technique,
the new set of Green functions are gauge independent at any order,  and provide a direct
way to evaluate form-factors, effective
charges, resonant transition amplitudes and the dynamically generated gluon mass\cite{binosi,aguilar}.

Unfortunately, there is no formal functional definition of the procedure which is {\it operational}
and depends on the given diagrammatic expansion of the theory.
However, the existence of a direct correspondence between pinch technique and background-field
method has led to a new set of modified SD equations which are satisfied by the new gauge-independent
Green functions. The modified SD equations turn out to be the 
SD background-field equations in the Feynman gauge\cite{BMPT}.
Thus, operationally, we can define the new gauge-independent functions as the solutions of the background-field
SD equations, provided that we set $\xi=1$.

For instance, the effective gluon propagator $\widehat{\Delta}(p)$ is a {\it physical} function
endowed with interesting features and directly related to an effective  renormalized coupling\cite{binosi}.
Thus, it would be interesting to evaluate that function by an analytical method like the
one-loop screened expansion, in order to study the analytic properties in the complex plane.
Actually, the function $\widehat{\Delta}(p)$ is tightly linked to the dressed propagator $\Delta(p)$
and we argue that the two functions might share the same poles.

Having derived the screened expansion from the SD equations, we can easily modify the expansion and 
write a screened expansion for the gauge-independent $\widehat{\Delta}(p)$ starting from the
background-field SD equations in the Feynman gauge.
The new one-loop expansion defines an approximate analytical solution of the equations which would be
equivalent to an untrivial non-perturbative truncation of the exact equations. As it happens for the standard 
gluon propagator $\Delta(p)$, we expect that a variationally improved expansion would lead to a very accurate approximation for the effective gauge-independent function $\widehat{\Delta}(p)$.

On the other hand, the modified SD equations have a structure similar to the usual SD set in Eq.(\ref{SDE}).
At one loop, the graphs contributing to the gluon self-energy are exactly the same, but the structure of the
bare vertices is modified because all external gluon lines must be regarded as background gluons. 
Then, the screened expansion would lead to the same one-loop graphs
reported in Fig.~2, but the resulting analytical expressions are different of course, because the background-field vertices must be used instead of the usual ones.

In more detail, in the background-field method, 
the structure of the SD equations is the following\cite{binosi}
\begin{align}
&\widehat{\Delta}^{-1}=\Delta_0^{-1}-\widehat{\Pi}\nn\\
&\widehat{\Pi}=\widehat{\Pi}\big[\Delta,D,\Gamma_i\big]\nn\\
&\Delta=\left(1+G\right)^2\widehat{\Delta}^{-1}\nn\\
&D^{-1}=D_0^{-1}-\Sigma\nn\\
&\Sigma=\Sigma\big[\Delta,D,\Gamma_i\big]\nn\\
&\Gamma_i=\Gamma_i\big[\Delta, D,\Gamma_j\big]\nn\\
&G=G\big[\Delta, D,\Gamma_j\big].
\end{align}
where the new scalar function $G$ relates the effective gluon propagator $\widehat{\Delta}$ to the
transverse part of the gluon propagator $\Delta$ which  enters on the right-hand side of the equations.
Here, the  conventions are the same as in Eq.(\ref{SDE}), with the exact longitudinal parts of the full propagators
given by Eq.(\ref{Delta}). Moreover, the function $G$ has no tree-level contribution and its leading term is
of order $g^2$, so that at one-loop, it can be set to zero  inside the loops, where $\Delta$ can be replaced
by $\widehat{\Delta}$. The one-loop graphs of the functional $\widehat{\Pi}$ are the same one-loop graphs of 
Fig.~1 but all the external gluon lines must be regarded as background fields\cite{aguilar, binosi}.

At one loop, the screened expansion can be easily obtained as discussed in the previous section.
For $N=1$, in the {\it minimal} approach, the one-loop SD equations follow by the same steps that led to 
Eq.(\ref{SDE1st1L}), yielding
\begin{align}
&\widehat{\Delta}^{-1}=\Delta_m^{-1}-\widehat{\Pi}^\prime\nn\\
&\widehat{\Pi}^\prime=m^2+\left(1-m^2\frac{\partial}{\partial m^2}\right)
\widehat{\Pi}^{(1L)}\big[\Delta_m, D_0,\Gamma_i^{(0)}\big]\nn\\
&D^{-1}=D_0^{-1}-\Sigma\nn\\
&\Sigma=\left(1-m^2\frac{\partial}{\partial m^2}\right)
\Sigma^{(1L)}\big[\Delta_m,D_0,\Gamma_i^{(0)}\big]\nn\\
&\Gamma_i=\left(1-m^2\frac{\partial}{\partial m^2}\right)\Gamma_i^{(1L)}\big[\Delta_m, D_0,\Gamma_j^{(0)}\big]
\label{SDEPT}
\end{align}
and of course, the explicit graphs contributing to $\widehat{\Pi}^\prime$ and $\Sigma$ are the same graphs of
Fig.~2, with the gluon vertices replaced by the corresponding background-field ones in the effective gluon
self-energy $\widehat{\Pi}^\prime$. All graphs must be evaluated in the Feynman gauge, $\xi=1$, in order
to obtain an approximation for the gauge-parameter-independent effective gluon propagator $\widehat{\Delta}$.

In the {\it democratic} vertex-counting scheme, the doubly crossed tadpole, which is shown in braces in Fig.~2, must be included, even if it has $N=2$, since it is a third-order one-loop graph, containing three vertices. While the relevance of this finite term might be questioned for the gluon propagator $\Delta$,
here its presence is crucial for the effective propagator $\widehat{\Delta}$. As previously discussed,
the expansion in Eq.(\ref{resum}) must be truncated for inserting some non-perturbative content, but
the truncation order is quite arbitrary, especially when the omitted terms are finite. Thus,
the pinch-technique provides an interesting motivation for retaining the doubly crossed tadpole, 
since without it, the one-loop
effective self energy $\widehat{\Pi}$ would be exactly zero at $p=0$, invalidating the strict
$\lambda$-expansion in powers of  $\Pi^\prime$ which would be of order $m^2$, 
according to Eqs.(\ref{SDprime}),(\ref{lambda}).
On the other hand, the doubly crossed tadpole appears as the first non-vanishing contribution to
$\widehat{\Pi}$ at $p=0$ and, in the vertex-counting scheme, its natural inclusion  restores Eq.(\ref{lambda}).
Then, besides its physical relevance, the effective propagator $\widehat{\Pi}$ provides an interesting
example where the simple {\it minimal} expansion does not work and the more involved vertex-counting
scheme must be used instead.

It is instructive to go through the details of the calculation in this simple case and evaluate
the effective self-energy $\widehat{\Pi}$ at $p=0$.
At the strict order $N=L=1$, the accidental vanishing of $\widehat{\Pi}(0)$
is expected, as a consequence of the QED Ward identity which is satisfied by the modified vertices.
The spurious mass divergence is already canceled in the ordinary loops and then the crossed
graphs cancel the residual finite mass entirely at $N=1$, requiring the inclusion of higher order terms.

The three-gluon and four-gluon vertices involved, with one external background line, were reported 
in Ref.\cite{abbott81}. They read

\begin{align}
i\Gamma^{\alpha\mu\nu}_{abc}=g\,f^{abc} &[(p_2- p_1)_\alpha \>g_{\mu\nu}\nn\\
&+(p_1-q+\frac{1}{\xi} p_2)_\nu \>g_{\mu\alpha}\nn\\
&+(q-p_2-\frac{1}{\xi} p_1)_\mu \>g_{\nu\alpha}]
\end{align}

\begin{align}
i\Gamma^{\mu\nu\rho\sigma}_{abcd}&=
-ig^2[f^{xab} f^{xcd}[(g_{\mu\rho} g_{\sigma\nu}-g_{\mu\sigma} g_{\nu\rho})\nn\\
&+f^{xac} f^{xdb}[(g_{\mu\sigma} g_{\rho\nu}-g_{\mu\nu} g_{\rho\sigma}
-\frac{1}{\xi} g_{\mu\rho} g_{\sigma\nu})\nn\\
&+f^{xad} f^{xbc}[(g_{\mu\nu} g_{\sigma\rho}-g_{\mu\rho} g_{\sigma\nu}
+\frac{1}{\xi} g_{\mu\sigma} g_{\nu\rho}).
\end{align}

The gluon loop, graph (2b) in Fig.~2, can be evaluated by inserting the gluon propagator of
Eq.(\ref{Delta}) with $\Delta=\Delta_m$. 
Taking the external momentum $p=0$ and setting $\xi=1$, the contribution of the
longitudinal propagator is canceled by the vertex structure and we can write, in Euclidean space,
the transverse projection of the graph as
\begin{align}
\widehat{\Pi}^{(2b)}(0)=&2N_c (g\mu^\epsilon)^2\>\frac{(d-1)}{d}\int \kkEd\frac{k_E^2}{(k_E^2+m^2)^2}\nn\\
=&-\frac{3N_c g^2}{(4\pi)^2}\>m^2\>\left(\frac{1}{\epsilon}+\log\frac{\mu^2}{m^2}+\text{const.}\right).
\end{align}

The transverse projection of the uncrossed tadpole, graph (1b) in Fig.~2, follows as
\begin{align}
\widehat{\Pi}^{(1b)}(0)=&-N_c(g\mu^\epsilon)^2\>  (d-1)\int \kkEd\>\frac{1}{k_E^2+m^2}\nn\\
=&\frac{3N_c g^2}{(4\pi)^2}\>m^2\>\left(\frac{1}{\epsilon}+\log\frac{\mu^2}{m^2}+\text{const.}\right).
\end{align}

As expected, the spurious mass divergence is canceled and the sum is finite
\BE
\widehat{\Pi}^{(1b)}(0)+\widehat{\Pi}^{(2b)}(0)=\text{const.}\times m^2
\EE
but that constant term would depend on the regularization scheme.
However, even the spurious constant term disappears when the crossed graphs, (1c) and (2c) in Fig.~2, are added,
yielding the trivial result
\BE
\left(1-m^2\frac{\partial}{\partial m^2}\right)\> m^2=0
\EE
and $\widehat{\Pi} (0)=0$.  We are not considering the ghost loop which is zero in the limit
$p\to 0$.

We have not added yet the doubly crossed tadpole, graph (1d) in Fig.~2, which should
be included by the vertex-counting criterion. By a double derivative, as shown in Eq.(\ref{m4})
\begin{align}
\widehat{\Pi}^{(1d)}(0)&=\frac{m^4}{2}\frac{\partial^2}{\partial (m^2)^2}
 \widehat{\Pi}^{(1b)}(0)
=-\frac{3N_c g^2}{(4\pi)^2}\>m^2
\end{align}
which is finite, does not depend on any spurious constant and gives the total contribution to the
effective self-energy $\widehat{\Pi}(0)$, yielding
\BE
\widehat{\Pi}^\prime (0)=m^2-\frac{3N_c g^2}{(4\pi)^2}\>m^2.
\EE
Finally, by Eq.(\ref{SDEPT}), the effective propagator reads
\BE
\widehat{\Delta}^{-1}(0)=\Delta_m^{-1}(0)-\widehat{\Pi}^\prime(0)=\frac{3N_c g^2}{(4\pi)^2}\>m^2
\EE
so that $\widehat{\Delta}(0)$ is finite (and positive).
In fact, a renormalization-group invariant can be defined as\cite{BMPT}
\BE
\widehat{d}=g^2\widehat{\Delta}
\EE
and this object can be regarded as an effective coupling which saturates in the IR.

\section{Closing remarks and outlook}

The screened massive expansion is a useful perturbative tool for the study of QCD in the
non-perturbative regime and its two-loop extension would provide valuable information
on the analytic properties of the correlators and on related problems like
confinement, mass generation and vacuum elementary excitations.

Here, we formulated the expansion as a modified loop expansion of the exact  SD equations.
The new formulation provides a general scheme for extending the expansion to higher orders and
to different theories. Different truncation strategies are discussed and a minimal two-loop
expansion is discussed, which seems to be free of spurious diverging mass terms. Thus, even at two-loop,
the correlators can be renormalized by the standard set of counterterms, without adding
spurious parameters. Moreover, the explicit calculation of the graphs would follow by simple derivatives
of the massive graphs which appear in the well studied Curci-Ferrari model.

Even at one-loop, the present method is useful for discussing different truncation strategies.
A simple ``convergence'' principle would suggest to adopt a {\it minimal} truncation scheme, where
no further finite terms are retained, once the cancellation of all the spurious divergences is achieved.
However, a {\it democratic} vertex-counting scheme seems to be more effective at one-loop and
might become crucial in other frameworks, like the pinch-technique.

Actually, one of the merit of the present formulation is its ability to describe different theories,
and the pinch-technique provides an interesting example since the occurrence of a mass term
is  prohibited by the QED Ward identities at one-loop. While a mass generation becomes harder in
that framework, the screened expansion seems to be robust enough and predicts a finite effective gluon propagator in the IR, albeit in the more effective vertex-counting scheme.

It would be interesting to pursue the present study in the two open directions: a full two-loop calculation
of the QCD correlators and a one-loop analytical calculation of the effective gluon propagator by
the pinch-technique.

For instance, the poor one-loop description of the quark renormalization function improves
by a two-loop calculation in the massive Curci-Ferrari model\cite{pelaez21,2LQ}. 
It would be very interesting to
see if the same improvement can be achieved by the screened massive
expansion, which would provide a reliable precise calculation from first principles.

At variance with numerical solutions of the SD equations, a one-loop calculation of the effective
gluon propagator would give explicit analytical results which could be easily continued to the
whole complex plane. Since the effective propagator is a {\it physical} gauge-invariant object, the location
of its poles would be very important, to be compared with the complex-conjugated poles which
have been reported for the Yang-Mills propagator\cite{analyt,xigauge} and seem to be related to the gluon confinement\cite{damp,poles}.
Thus, a detailed study of the analytic properties of the effective propagator would be very welcome.

\acknowledgments

This research was supported in part by  ``Linea di intervento 2''  as project HQCDyn of
Dipartimento di Fisica e Astronomia, Universit\`a di Catania and by the national project SIM of
Istituto Nazionale di Fisica Nucleare.

\end{document}